\documentclass[twocolumn,prd,aps,groupedaddress,nopacs,nofootinbib,floats,floatfix]{revtex4}

\newcommand{\be}{\begin{equation}}
\newcommand{\ee}{\end{equation}}
\newcommand{\ba}{\begin{eqnarray}}
\newcommand{\ea}{\end{eqnarray}}
\renewcommand{\d}{\mathrm{d}}

\usepackage{color}

\usepackage{amsmath}
\usepackage{mathrsfs}
\usepackage{amssymb}

\numberwithin{equation}{section}

\usepackage{epsfig}

\newcommand{\hu}{kms$^{-1}$Mpc$^{-1}$}

\def\beq{\begin{equation}}
\def\eeq{\end{equation}}
\def\ber{\begin{eqnarray}}
\def\eer{\end{eqnarray}}

\def \lleq {\lower0.9ex\hbox{ $\buildrel < \over \sim$} ~}
\def \ggeq {\lower0.9ex\hbox{ $\buildrel > \over \sim$} ~}
\def\apj{{Astroph.\@ J.\ }}
\def\mn{{Mon.\@ Not.\@ Roy.\@ Ast.\@ Soc.\ }}
\def\asta{{Astron.\@ Astrophys.\ }}

\def\prl{{Phys.\@ Rev.\@ Lett.\ }}
\def\prd{{Phys.\@ Rev.\@ D\ }}

\def\etal{{\it et al.}}

\begin{document}

\title{Model independent tests of the standard cosmological model}

\author{Arman Shafieloo$^1$ \& Chris Clarkson$^2$\\
\it $^1$Department of Physics, University of Oxford, 1 Keble Road, Oxford, OX13NP, UK.\\
\it $^2$Centre for Astrophysics, Cosmology and Gravitation, and, Department Mathematics
and Applied Mathematics, University of Cape Town, Rondebosch 7701,
South Africa.}

\begin{abstract}

The dark energy problem has led to speculation that not only may $\Lambda$CDM be wrong, but that the FLRW models themselves may not even provide the correct family of background models. 
We discuss how direct measurements of $H(z)$ can be used to formulate tests of the standard paradigm in cosmology. On their own, such measurements can be used to test for deviations from flat $\Lambda$CDM. When combined with supernovae distances, Hubble rate measurements provide a test of the Copernican principle and the homogeneity assumption of the standard model, which is independent of dark energy or metric based theory of gravity. A modification of this test also provides a model independent observable for flatness which decorrelates curvature determination from dark energy. We investigate these tests using Hubble rate measurements from age data, as well as from a Hubble rate inferred from recent measurements of the baryon acoustic oscillations. While the current data is too weak to say anything significant, these tests are exciting prospects for the future.

\end{abstract}
\maketitle

\section{Introduction}

There is a growing realisation in cosmology that the dark energy problem means that we have to test the fundamentals of our cosmological model as rigorously as we can. Despite the fact that the standard flat $\Lambda$CDM paradigm works so well on many fronts, the unknowns at the heart of it have led us into the territory of trying to invert observations to ascertain directly any temporal evolution of dark energy~-- whether it is in fact modified gravity, global inhomogeneity such as a void model, quintessence and so on. Normally in physics, we would hope to work the other way round: a model based on known physics can predict outcomes of an experiment which can then potentially falsify the model (or theory). Unfortunately, once we move away from the comfort of $\Lambda$CDM we have little physics to guide us at all; worse, many suggestions appear degenerate or even unfalsifiable. Generalised gravity theories, quintessence models, void models, etc. all have free \emph{functional} degrees of freedom, and not just free parameters.  So dark energy reconstruction tries to deduce functional degrees of freedom directly from data sets we have.  

In particular, if Hubble-scale inhomogeneity is responsible for the apparent acceleration, then all bets are off as to what the effective equation of state\footnote{That is, the equation of state which reproduces some particular family of observations, such as the distance modulus, if we utilise a FLRW model.} could be. In the simplest void models which describe this~\cite{voids,FLSC}, a violation of the Copernican principle~-- that we are not located at a special place in the universe~-- is required, because they place us very near the centre of symmetry. Clearly such a radical suggestion warrants much more study; for example, it is not yet understood how perturbations evolve, although the generalised Bardeen equation is known~\cite{CCF}.

One important aspect of the reconstruction approach will rely on formulating consistency tests of the standard paradigm. Ideally these tests would be constructed so that signals of deviations from (flat) $\Lambda$CDM are easy to spot and quantify. If it's possible, we would like these tests to be independent of the standard paradigm, and rely as little as possible on assumptions that we normally take for granted. For example, consistency between the background evolution and the evolution of perturbations can be used to rule out whole classes of quintessence models~\cite{MHH}. Using perturbations in this way is only possible within given classes of dark energy, however, because of assumptions about the sound speed which must be made, amongst other things. Tests which may be formulated purely from background observables would not suffer this difficulty.

Here our principle aim is to investigate the curvature test, introduced in~\cite{CBL}. This is a test which uses background observables only, and tests for consistency of the FLRW paradigm itself. In FLRW models, the luminosity distance may be written as
\begin{equation}\label{d_L}
d_{L}(z)=\frac{c(1+z)}{H_0 \sqrt{-\Omega_k}}\sin{\left( 
\sqrt{-\Omega_k}\int_0^z{\mathrm{d}z'\frac{H_0}{H(z')}}\right)},
\end{equation}
where $\Omega_k$ is the curvature parameter today, and the expansion rate $H(z)$ takes the value $H_0=H(0)$ today. (Note that this is actually valid for either sign of curvature; for $\Omega_k=0$ we can take the limit $\Omega_k\to0$.) The area distance is defined using $d_L=(1+z)^2d_A$, and another distance measure we will use is $D=(1+z)d_A/(c/H_0)$, which is the dimensionless comoving distance. We may rearrange Eq.~(\ref{d_L}) to give an expression for the curvature parameter in terms of $H(z)$ and $D(z)$~\cite{CBL}:
\begin{equation}\label{OK}
\Omega_k=\frac{\left[h(z)D'(z)\right]^2-1}{[D(z)]^2}=\mathscr{O}_k(z),
\end{equation}
where $h(z)=H(z)/H_0$ and $'=\d/\d z$.
This tells us how to measure the value of the curvature parameter today from distance and Hubble rate observations, independently of any other model parameters or dark energy model or field equations. Remarkably this tells us the curvature today from these measurements at any single redshift, provided that $D(z)$ and $H(z)$ come from unrelated families of observations. 
If the underlying geometry of the background cosmology is FLRW, $\mathscr{O}_k(z)$ will be measured as constant and so be independent of the redshift of measurement. As a consistency test of the standard paradigm it has all the right ingredients. It is a straightforward comparison of different observables, both of which can be formulated fairly independently of contamination from the standard model.  At the simplest level, then, we can use this to see if our distance measurements are consistent with our $H(z)$ measurements, if we choose to stick within the standard paradigm. But we can get more information than that.

For measurements of $\mathscr{O}_k(z)$  not to be constant as a function of redshift we are looking at pretty radical changes to the standard model~-- an alteration to the underlying homogeneous and isotropic FLRW models themselves. One important situation where this will happen is in void or Hubble-bubble models. These models fit the SNIa distance modulus by having us near the centre of a large Hubble-scale depression in the energy density and accompanying curvature.  These models violate the Copernican principle which lies at the heart of cosmology. If isotropy about us were determined at all redshifts, this curvature test may then be used as a direct test of the Copernican principle.  However, deviations from $\mathscr{O}_k(z)=$const. will be signalled in any non-FLRW cosmology; in particular, in Swiss-cheese models which satisfy the Copernican principle on very large scales will have an $\mathscr{O}_k(z)$ which oscillates with redshift. Furthermore, backreaction and averaging will induce non-constant $\mathscr{O}_k(z)$~\cite{wilt}.  (Note that in all these models it won't necessarily be saying anything specific about the curvature.)

\begin{figure}[htbp]
\begin{center}
\includegraphics[width=0.95\columnwidth]{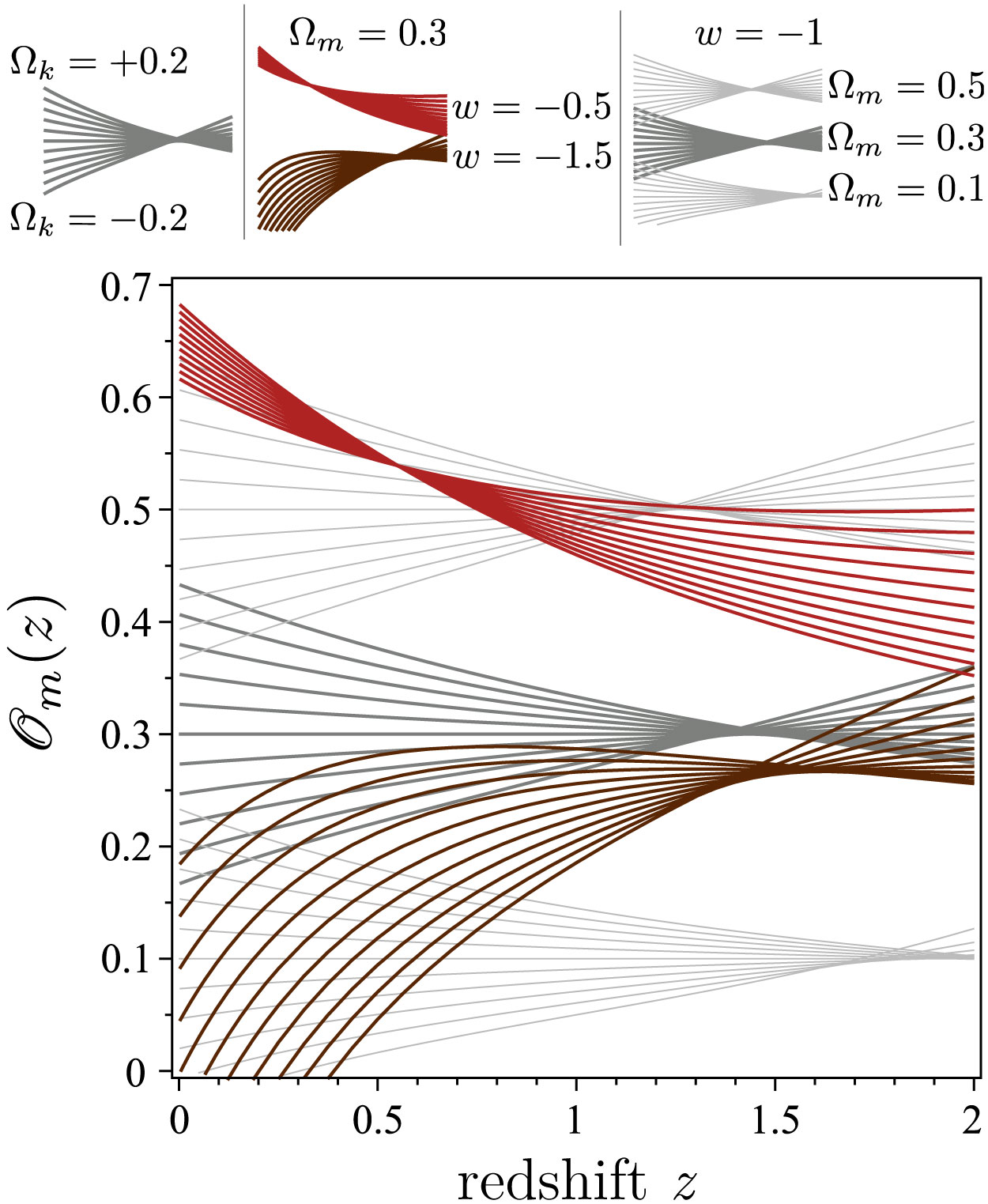}
\caption{A plot of $Om(z)$ in different models, obtained using distances. The figure shows the range of behaviour from curvature in each fan of curves (key, top left). The three grey fans show how differing $\Omega_m$ values interact with curvature for $\Lambda$CDM (key, top right). The effect of changing $w$ by a constant is illustrated in the red and brown fans. Similar curves are found if we use $h$ rather than $D$ to form $Om(z)$, but without the cross-over at high redshift. It is clear that, although $Om(z)$ deviates from a constant because of curvature this deviation is not strong for small curvature values; therefore, $Om(z)$'s utility lies in picking up non-$\Lambda$ behaviour.}
\label{fig:om}
\end{center}
\end{figure}

A similar background test which is specifically to test any deviations the flat $\Lambda$CDM scenario was introduced independently in~\cite{om,litmus}. The $Om$ diagnostic defined by~\cite{om}:
\be
\label{eq:om}
Om(z)=\frac{h^2(z)-1}{(1+z)^3-1}=\frac{1-D'(z)^2}{[(1+z)^3-1]D'(z)^2},
\ee
is a constant at different redshifts in a spatially flat $\Lambda$CDM model, equal to today's value of the matter density parameter $\Omega_m$. In a similar spirit to $\Omega_k$ we can measure the variables on the rhs, and expect our measurements to yield the same number regardless of redshift. Fig.~\ref{fig:om} shows the kind of behaviour $Om(z)$ exhibits from different FLRW models.  
From this we can also derive~\cite{litmus}:
\ba
\label{eq:L}
\mathscr{L}(z)\!\!&=&\!\!2[(1+z)^3-1] D''(z)\nonumber\\&&
+3(1+z)^2D'(z)[1-D'(z)^2]\nonumber\\
&=&\!\!0~\mathrm{for~all~flat}~\Lambda\mathrm{CDM~models}.
\ea
Testing for deviations from $\mathscr{L}(z)=0$ provides an alternate test of flat $\Lambda$CDM.

The main issue in using the curvature test lies in observing $H(z)$ independently of distance measurements. For example, $H(z)$ is often reconstructed from  supernovae data~\cite{varun_alexei_rev06}, but this implicitly uses Eq.~(\ref{OK}) and so can't be used here. There are other methods, two of which we use here. Measurements of the Baryon Acoustic Oscillations give the `volume distance' which is~
\be\label{DV}
D_V(z)^3=\left(\frac{c}{H_0}\right)^3\frac{z D(z)^2}{h(z)},
\ee
from which we can derive $H(z)$ if we have an another method to determine $D(z)$. Alternatively, we can reformulate the curvature test directly in terms of $D_V(z)$ and $D(z)$. This method is not entirely independent of cosmological model, however, because it assumes that there is no scale dependence in the evolution of perturbations~-- this is not the case in inhomogeneous models where the curvature is varying spatially~\cite{CCF}.  

Spectroscopic age dating of passively evolving galaxies give another measure of $H(z)$~\cite{jimenez_verde}, which we use here. This method measures 
\be
\frac{\d t}{\d z}=\frac{1}{(1+z)H(z)}.
\ee 
Other possible methods include measurements of the time drift of redshifts~\cite{zdot} and the dipole in the SN observations~\cite{ruth}, but these are not considered further here. 
For distance measurements we shall use the latest complete SNIa data set~\cite{Hicken09}. By using the smoothing method presented in~\cite{Shafieloo05,Shafieloo07,Shafieloo09b} we can turn this data into $D(z)$ and $D'(z)$ without relying on any fiducial cosmological model or parameterisation. 
We use $\mathscr{O}_k(z) = \Omega_k$ in Eq.~(\ref{OK}) and $Om(z)$  in Eq.~(\ref{eq:om}) as  two diagnostics 
which we attempt to reconstruct using these independent datasets. Any deviations from constant will signify problems with the FLRW models themselves in the case of $\mathscr{O}_k(z)$, and problems with flat $\Lambda$CDM in the case of $Om(z)$. Finally we make use of a third test: $\mathscr{O}_k(z)D(z)^2$ may be used as a model independent test for flatness if we look for deviations from zero.

\section{Construction of the diagnostics}

\subsection{Finding $D(z)$ and $D'(z)$ from supernovae}

To find $D(z)$ and $D'(z)$, we use the smoothing method implemented and applied to supernovae data by \cite{Shafieloo05,Shafieloo07}. In addition we have done some modifications to the smoothing method to make it error-sensitive that results to a better fit to the data~\cite{Shafieloo09b}. 
This method is a completely model independent approach to derive the $d_L(z)$ relation directly from the data, without any assumptions other than the introduction of a smoothing scale. The only parameter used in the smoothing method is the smoothing width $\Delta$, which is constrained only by the quality and quantity of the data, and has nothing to do with any cosmological model. The smoothing method is an iterative procedure with each iteration typically giving a better fit to the data. It has been shown in \cite{Shafieloo05,Shafieloo07} that the final reconstructed results are independent of the assumed initial guess, $d_L(z_i)^g$ below. 

The modified smoothing method (error-sensitive) can be summarized by the following equation~\cite{Shafieloo09b}:
\ber
\label{eq:bg}
&&\ln d_L(z,\Delta)^{\rm s}=\ln
\ d_L(z)^g\nonumber\\&& +N(z) \sum_i \frac{\left [ \ln d_L(z_i)- \ln
\ d_L(z_i)^g \right]}{\sigma^2_{d_L(z_i)}} 
 \ {\rm exp} \left [- \frac{\ln^2 \left
( \frac{1+z_i}{1+z} \right ) }{2 \Delta^2} \right ],  \nonumber\\&&~~~\text{where}~\nonumber\\
&&N(z)^{-1}=\sum_i {\rm exp} \left
[- \frac{\ln^2 \left ( \frac{1+z_i}{1+z} \right ) }{2 \Delta^2} \right ] \frac{1}{\sigma^2_{d_L(z_i)}} ~. 
\eer
where $d_L(z)$ is the data, $N(z)$ is the normalization factor, $d_L(z_i)^g$ is the initial guess model and $\Delta$ is the width of smoothing.

\begin{figure}[!h]
\includegraphics[width=0.95\columnwidth]{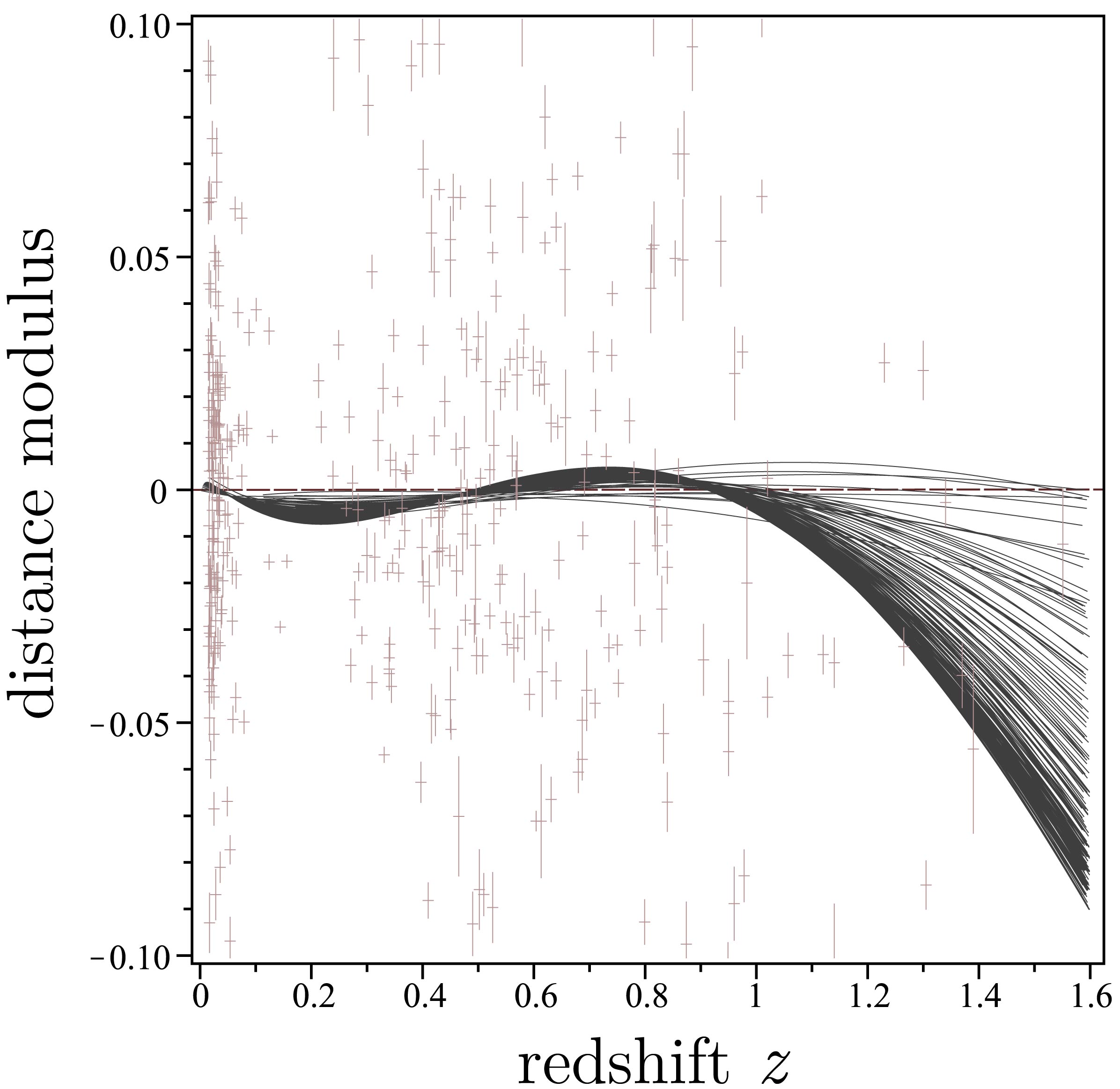}
\caption{The best fit curves to the constitution supernovae, shown here with vertical error bars reduced by a factor of 10 for clarity.  The distance modulus is shown with the best fit flat $\Lambda$CDM subtracted off (with $H_0=65$~\hu). The curves all have a better fit to the data than $\Lambda$CDM, shown by the dashed horizontal line. Note the lack of SNIa above $z\sim1$ means that $D(z)$ is not well constrained there.}
\label{fig:D}
\end{figure}

The absolute brightness of the supernovae is degenerate with $H_0$ since the observed quantity is the distance modulus $\mu(z)$. The outcome of the smoothing method is therefore $H_0d_L(z)/c\equiv d_L^\text{rec}(z)=(1+z)D(z)$. We should note that the actual value of $H_0$ is quite important in our analysis; supernovae surveys usually assume the value of $H_0$ in their data which can bias our results. In this paper we use the recent `constitution' supernovae data~\cite{Hicken09} that include 397 data points and we choose $\Delta=0.30$. This value of $\Delta$ is half of the value used in \cite{Shafieloo07} in analysis of the SNLS data \cite{SNLS}, simply because we have almost four times as many data points in the constitution sample in comparison with SNLS data. Complete explanation of the relations between the $\Delta$, the number of data points, quality of the data and the reconstructed results can be found in~\cite{Shafieloo05,Shafieloo07}. Our best reconstructed result using smoothing method gives $\chi^2=459.4$ to the constitution data, which is a significantly better fit than the best fit $\chi^2$ derived by using the CPL parameterization $w(z)=w_0+w_1\frac{z}{1+z}$~\cite{CPL}, with $\Delta \chi^2 \approx 1.7$ \cite{Shafieloo09a}. By using our smoothing method we reconstruct a set of $d_L^\text{rec}(z)$ in the redshift range of the supernovae data. We consider only the reconstructed results that have a better fit than the best fit spatially flat $\Lambda$CDM model with $\chi^2=465.63$ ($\Omega_{0m}=0.288$)~\cite{Shafieloo09a}. We show these curves in Fig.~\ref{fig:D}. {We should note that though all curves we have shown in Fig.~\ref{fig:D} have the $\chi^2$ better than the best fit $\Lambda$CDM model, but we should consider that the $\Delta \chi^2$ between these reconstructions of the expansion history and the best fit $\Lambda$CDM model is not that large to conclude that flat $\Lambda$CDM has a strong discrepancy with the data (maximum $\Delta \chi^2$ is around 6).}

Once we have reconstructed $D(z)$ finding $D'(z)$ is a matter of differentiating our $D(z)$ curves. Because the procedure is iterative the errors on $D'(z)$ may be estimated from the set of $D(z)$ curves which we consider with $\chi^2 < 465.63 $. 

\begin{figure*}[ht!]
\includegraphics[width=0.8\textwidth]{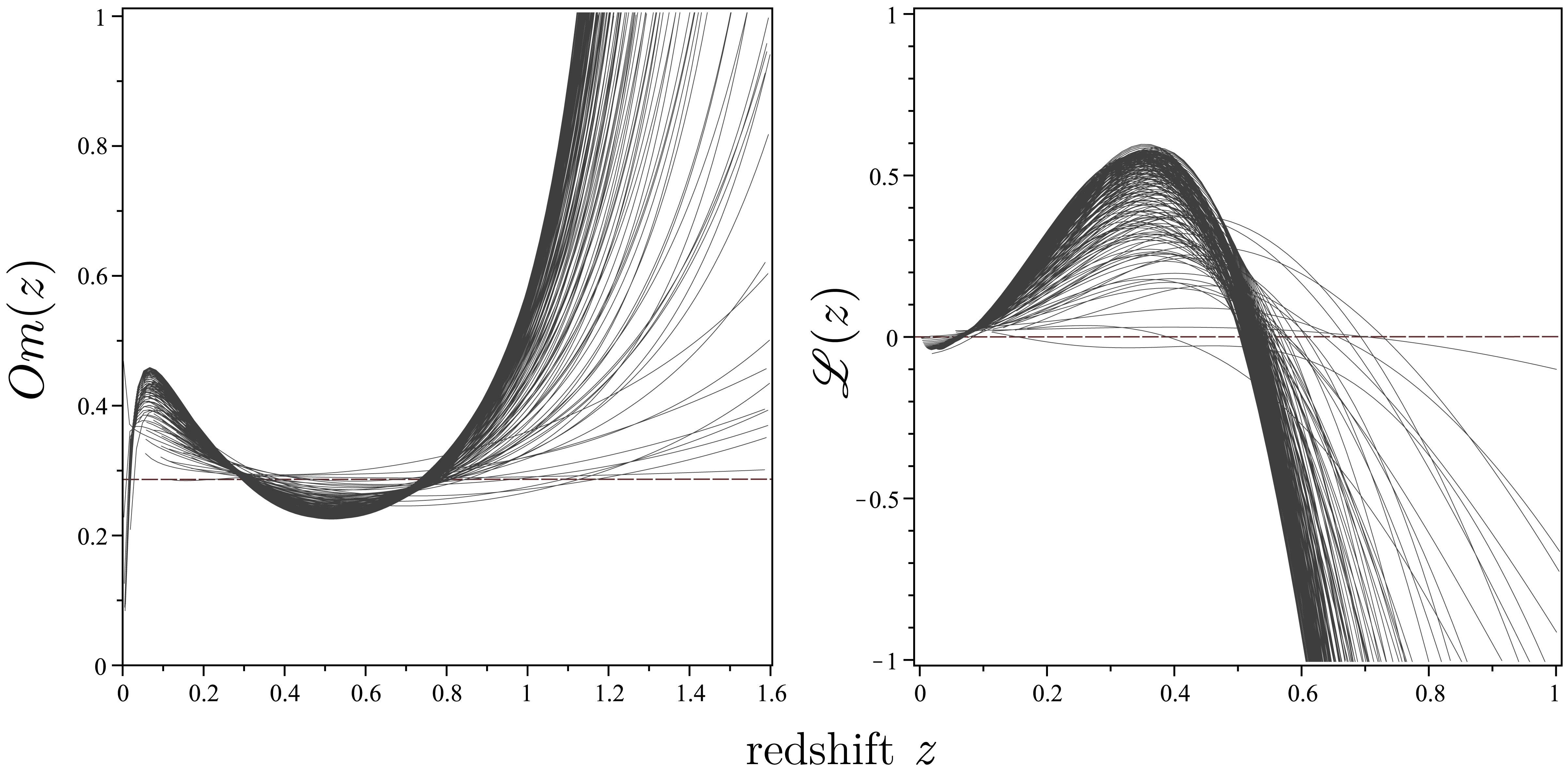}
\caption{Tests for flat $\Lambda$CDM, $Om(z)$ and $\mathscr{L}(z)$. The smoothing method allows us to construct these directly. The curves shown are all a better fit to the data than flat $\Lambda$CDM, which is shown as the dashed line in the plots. { These two plots are made from the same reconstruction.  We have shown $\mathscr{L}(z)$ up to $z=1$ because the error-bars at the higher redshifts are so large. $\mathscr{L}(z)$ is derived by using the second derivative of the results from the smoothing method were $Om(z)$ is using the first derivative, so the error-bars for $\mathscr{L}(z)$ are larger.}
}
\label{fig:omandL}
\end{figure*}

\subsection{Finding $h(z)$ from the BAO}

\subsubsection{$D_V(z)$ from the BAO and CMB}

From direct observations, there are two CMB parameters, $R$, the scaled distance to recombination, and $\ell_a$, the angular scale of the sound horizon at recombination, with measured values that are nearly uncorrelated with each other~\cite{Wang_Mukherjee_07}. These parameters are defined as:
\be 
R=\sqrt{\Omega_m H_0^2}r(z_{CMB})
\ee
and
\be
\ell_a = \frac{\pi r(z_{CMB})}{r_s(z_{CMB})},
\ee
where $r(z)=cD(z)/H_0$ is the comoving distance, and $z_{CMB}=1090$. 
It is also possible to put constraints on $r_s(z_{CMB})$, the comoving sound horizon at recombination, directly from CMB data independently of the assumption of the cosmic curvature or model of dark energy \cite{Wang_Mukherjee_07}:
\be
r_s(z_{CMB})=\frac{c}{H_0}\int_{z_{CMB}}^{\infty} dz\frac{c_s}{h(z)},
\ee   
where $c_s$ is the sound speed and the derived value of $r_s(z_{CMB})$ from WMAP data is $r_s(z_{CMB})=148.55 \pm 2.60$Mpc \cite{Wang_Mukherjee_07}.

The BAO give a measurement of $\mathscr{R}(z)=\frac{r_s(z_{CMB})}{D_V(z)}$ at two redshifts of $z=0.35$ and $z=0.20$~\cite{percival09}: $\frac{r_s(z_{CMB})}{D_V(0.20)}=0.1905 \pm 0.0061$ and $\frac{r_s(z_{CMB})}{D_V(0.35)}=0.1097 \pm 0.0036$. Combining these measurements with the value of  $r_s(z_{CMB})=148.55 \pm 2.60 Mpc$ from WMAP data we can estimate the values of $D_V(0.20)$ and $D_V(0.35)$. Since we are dealing with two completely different observations we can assume that they are almost uncorrelated. Hence we can derive $\sigma_{D_V(z)}$:
\be
{\sigma_{D_V(z)}}^2= \left[\frac{\partial D_V(z)}{\partial r_s(z_{CMB})}\right]^2{\sigma_{r_s(z_{CMB})}}^2+\left[\frac{\partial D_V(z)}{\partial \mathscr{R}(z)}\right]^2 {\sigma_{\mathscr{R}(z)}}^2
\ee
which results in $D_V(0.20)= 779.79 \pm 28.45$Mpc and $D_V(0.35)=1354.14 \pm 50.36$Mpc.

\subsubsection{Derivation of h(z)} 

From $D_V$ and $d_L^\text{rec}$ we may estimate $h(z)$ from
\be
h(z)=\left(\frac{c}{H_0}\right)^3\frac{zd_L^\text{rec}(z)^2}{(1+z)^2D_V(z)^3}.
\ee
To calculate $h(z)$, therefore, we also need to know the value of $H_0$. {The uncertainties of $H_0$ can affect the reconstruction of the cosmological quantities. As we see in eqs. 1.2 and 1.3, to test cosmological models we need to know $h(z)=H(z)/H_0$. So only knowledge of $H(z)$ is not enough and the value and error-bars of $H_0$ are also very important and they affect the reconstructed $h(z)$ and its error-bars. In fact the first term in the right hand side of eq. 2.7, is the error from the uncertainties of $H_0$.} For our analysis we choose different constraints of $H_0$ from different observations. For each reconstructed set of $d_L^\text{rec}(z)$ from supernovae data we can derive $\sigma_{h(z)}$:
\ba
{\sigma_{h(z)}}^2&=& \left[\frac{\partial h(z)}{\partial H_0}\right]^2{\sigma_{H_0}}^2+\left[\frac{\partial h(z)}{\partial D_V(z)}\right]^2 {\sigma_{D_V(z)}}^2
\nonumber\\
&=&\left[ 3 h(z) \right]^2\left\{\left[\frac{\sigma_{H_0}}{H_0}\right]^2+\left[\frac{\sigma_{D_V(z)}}{D_V(z)}\right]^2\right\}.
\ea
From $\sigma_{h(z)}$, we can evaluate $\sigma_{\mathscr{O}_k}$ and $\sigma_{Om}$:
\be\label{eq:sigok}
{\sigma_{\mathscr{O}_k(z)}}^2=\left[\frac{2{d_L^\text{rec}}'(z)^2h(z)}{d_L^\text{rec}(z)^2}\right]^2 \sigma_{h(z)}^2
\ee
and 
\be\label{eq:sigom}
{\sigma_{Om(z)}}^2=\left[\frac{2h(z)}{(1+z)^3-1}\right]^2 \sigma_{h(z)}^2 .
\ee
Note that these errors are for a given $d_L^\text{rec}(z)$; errors resulting from the SN Ia measurements are found by calculating $h(z)$ for many $d_L^\text{rec}(z)$.

\subsection{Finding $h(z)$ from age data}

In \cite{jimenez_verde,stern09} $H(z)$ is derived by using the relative ages of passively evolving galaxies. We can simply derive $h(z)$ and $\sigma^2_{h(z)}$ using this data:
\be
\sigma^2_{h(z)}={H_0}^{-2}\sigma_{H(z)}^2+h(z)^2{H_0}^{-2}\sigma_{H_0}^2.
\ee
This assumes that our measurements of $H_0$ and $H(z)$ are uncorrelated. 
We can now estimate $Om(z)$ and $\mathscr{O}_k(z)$ in a same way as we did in the previous section by using $D(z)$ and $D'(z)$ from supernovae data.

\section{Reconstructed Diagnostics}

\subsection{$Om(z)$ and $\mathscr{L}(z)$ from supernovae distances}

The tests for flat $\Lambda$CDM discussed above may be utilised directly using distance data and its derivatives, which we show in Fig.~\ref{fig:omandL}. 
Both $Om(z)$ and $\mathscr{L}(z)$ tell us useful information about possible deviations from flat $\Lambda$CDM. For example, at $z\approx0.3$, $Om(z)$ appears consistent with $\Omega_m\approx0.3$; $\mathscr{L}(z)$ tells a different story as it deviates from zero there.  For both curves there is a preferential  deviation from the concordance model. While increasing behaviour of $Om(z)$ at low redshifts suggests larger $w(z)$ for dark energy in this range, the large errors at high red shifts above $z \sim 1$ indicates poor quality of the data. While these curves are preferred over flat $\Lambda$CDM, they are not strongly preferred enough to get too excited about. It has been discussed before that even by using parametric methods in the reconstruction of the expansion history, the $Om$ diagnostic would not be very sensitive to the choice of parameterization; as we see here our results for $Om(z)$ using a non-parametric smoothing method are very similar to reconstructed $Om(z)$ using a series ansatz for $w(z)$~\cite{Shafieloo09a}. It is also rather surprising that the curves derived here, which are model independent constructions of $Om(z)$, have similar features to the corresponding curves we find in void models with a smooth centre~\cite{FLSC}.

\subsection{The curvature test and $Om(z)$ using Hubble rate measurements}

The diagnostics are sensitive to the value of $H_0$. Assuming free priors on curvature and also equation of state of dark energy, Ref.~\cite{Riess09} find $H_0=68.7 \pm 2.0$ \hu\ from a combination of WMAP5+BAO+high-$z$ SNe data. The most recent constraints on $H_0$ from the observation of 240 long period Cepheids by Ref.~\cite{Riess09} give us instead $H_0=74.2 \pm 3.6$ \hu. The latter of these results is more suitable for us because it makes fewer assumptions on the underlying cosmology; we will consider $\mathscr{O}_k(z)$ and $Om(z)$ using both of these measurements however.

\begin{figure}[ht]
\includegraphics[width=0.9\columnwidth]{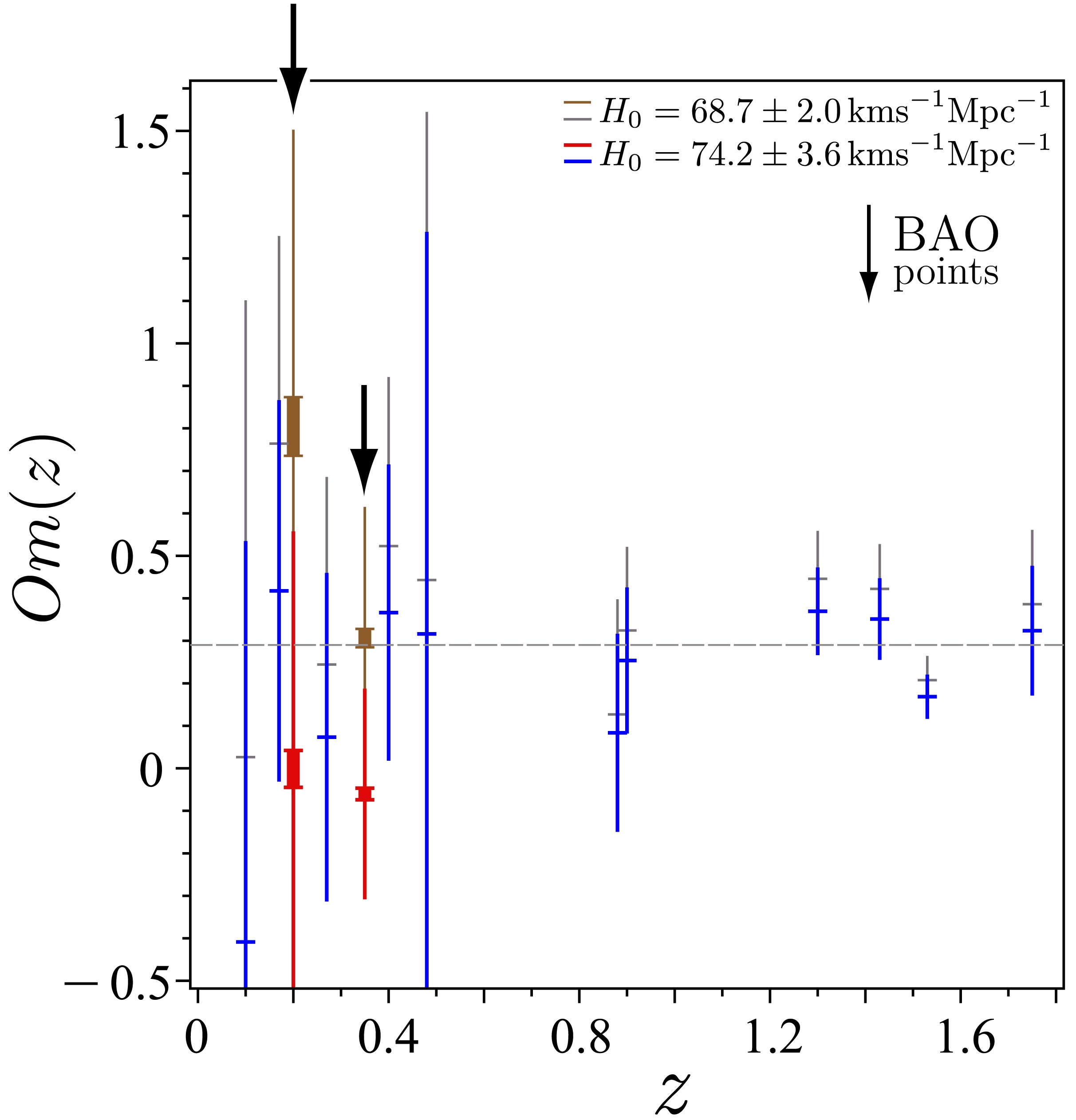}
\caption{
Resultant $Om(z)$ using two values of $H_0$.  The points indicated with an arrow show the SN (constitution) data combined with the recent BAO (SDSS-DR7) and CMB (WMAP)  data. The other points show the same using age data. The thick vertical lines represent the spread in the constructed point arising from uncertainty in the reconstructed distances; the thin vertical lines represent the errors arising from $H(z)$ and $H_0$. 
}
\label{fig:OM}
\end{figure}
\begin{figure*}[ht]
\includegraphics[width=0.98\textwidth]{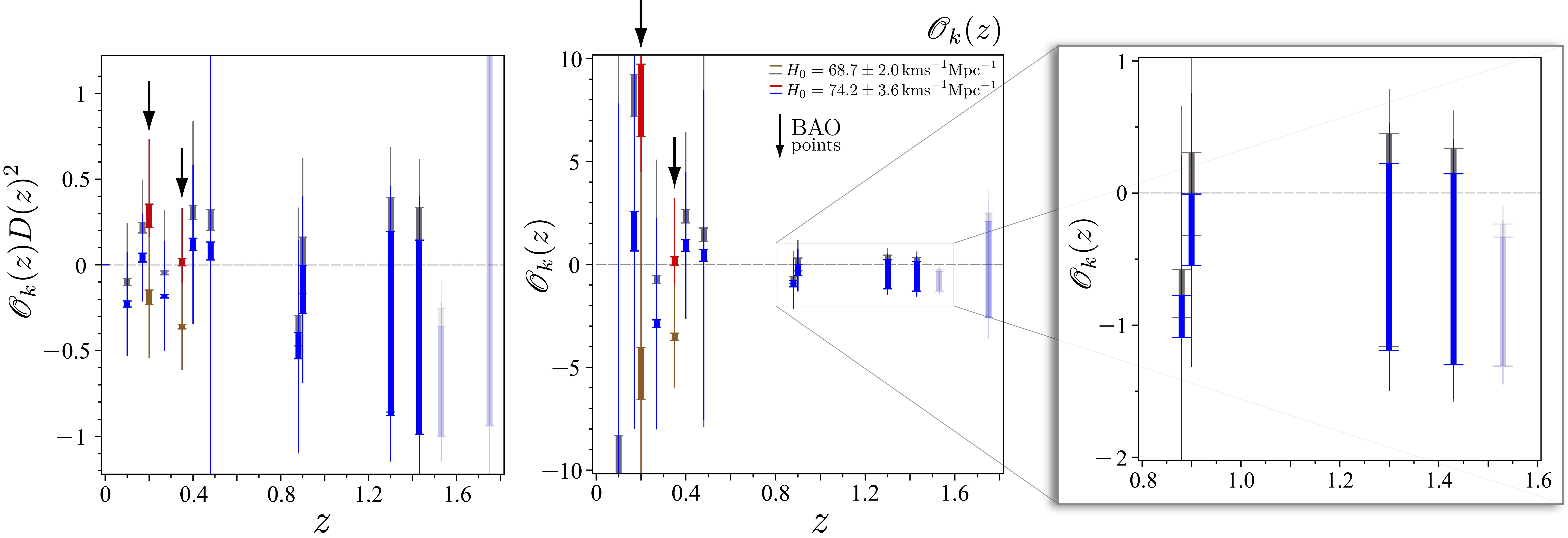}
\caption{
Resultant $\mathscr{O}_k(z)$ using two values of $H_0$.  The points indicated with an arrow show the SN (constitution) data combined with the recent BAO (SDSS-DR7) and CMB (WMAP)  data. The other points show the same using age data. The thick vertical lines represent the spread in the constructed point arising from uncertainty in the reconstructed distances; the thin vertical lines represent the errors arising from $H(z)$ and $H_0$. The faded points are two age data points which are at such high redshift they have no corresponding SNIa, so should not be taken too seriously. }
\label{fig:OK}
\end{figure*}

In Fig.~\ref{fig:OM} we present the $Om(z)$ diagnostic for $h(z)$ data from the BAO and age data (note that age data doesn't use the SN distances, but that the BAO does, when we reconstruct $h(z)$). These data points are essentially a remapping of the data through the function $Om(z)$. Note that the errors shrink with distance, and that the data are clearly consistent with $Om(z)=$const.

In Fig.~\ref{fig:OK}, we show the reconstructed $\mathscr{O}_k(z)$ using both BAO and age data to evaluate $h(z)$ using both values of $H_0$ (centre). For low redshift the errors are very large indeed, but for high redshift this test shows considerable promise. In the zoomed picture on the right, we can see that $\mathscr{O}_k(z)$, while not tightly constrained by current data, the error bars are reasonable in comparison with $Om(z)$, as most of the extra uncertainty comes from errors in the reconstructed $D'(z)$. We can see by eye that $\mathscr{O}_k(z)=$const. is consistent with the data, but that negative values are slightly favoured. This is actually  reflection of the fact that $Om(z)$ deviates strongly from constant at high redshift, which in turn reflects the fact that the SNIa data are not sufficient there.

The errors at low redshift indicate that $\mathscr{O}_k(z)=$const. will not be a useful test in this regime. However, the numerator of $\mathscr{O}_k(z)$, gives an important measure of deviations from flatness, irrespective of the dark energy equation of state or theory of gravity (see also~\cite{wilt}). In the left panel of Fig.~\ref{fig:OK} we see that for moderate redshift data points we have the seed of a model independent test for flatness which will become a robust test as the data improves. In general, measuring $\text{sign}(\mathscr{O}_k(z)D(z)^2)$ will give a robust and model independent determination of the sign of the spatial curvature.

The reason for the inefficiency of the Copernican test at low redshift may be understood as follows. Given two families of independent observations, $\{D(z),H(z)\}$, say, we may reconstruct separate FLRW models with their own parameters from each family of observations, modulo all the usual degeneracies. We may think of $\mathscr{O}_k(z)$ as the mismatch between these reconstructed models. For small redshifts, this implies that, 
\ba
\mathscr{O}_k(z)&=& \frac{\Omega_{k}^{(D)}-\Omega_{k}^{(H)}-3\left[\Omega_{DE}^{(D)}w_0^{(D)}-\Omega_{DE}^{(H)}w_0^{(H)}\right]}{z}\nonumber\\
&& + \mathcal{O}(z^0),
\ea
where $\Omega_{k}^{(D)}$ is the curvature parameter today evaluated from distance measurements, etc. The next term in the series is a constant and gives the value of the curvature parameter today when the two observables give the same model. For small redshift then, any errors are divergent. This limits the utility of this particular diagnostic to moderate to large redshift, where we have seen it shows potential. More generally, it will be useful to use the fact that
\begin{equation}
\mathscr{C}(z)=1+h^2\left(DD''-D'^2\right)+hh'DD'
\end{equation}
must be zero in any FLRW model~\cite{CBL}. This doesn't diverge at low redshifts; rather we have
\be
\mathscr{C}(z)=\left[q_0^{(D)}-q_0^{(H)}\right]z+\mathcal{O}(z^2),
\ee
which is much more stable. This comes at the expense of requiring derivatives of Hubble data, which we can imagine being dealt with using the smoothing method we have used here for $D(z)$ once the data becomes good enough.

\section{Discussion and conclusions}

We have investigated some model independent consistency tests of the FLRW models using a mixture of the latest complete SNIa data and some Hubble rate data. The SNIa data is turned into the dimensionless distance function $D(z)$ using a model independent smoothing method.   One source of the $H(z)$ data is from the BAO measurements of $D_V(z)$, which relies on assumptions about how perturbations evolve in the standard FLRW paradigm. The other is from the relative ages of passively evolving galaxies, which has far fewer assumptions about the underlying cosmological model built in to it. 

We have considered three tests:
\begin{itemize}
\item Tests for flat $\Lambda$CDM: $Om(z)$ and $\mathscr{L}(z)$. The first of these was investigated using both Hubble rate data and distance data independently. When using Hubble rate data, everything is consistent with flat $\Lambda$CDM. Using distance data, there seems to be slight evidence for deviation from flat $\Lambda$CDM. We should note that a larger $Om(z)$ at low red shifts, as we see in Fig.~\ref{fig:OM}, is indicative of a larger $w(z)$ for dark energy (considering the effect to be from dark energy, and not due to curvature) and suggest that some decaying models of dark energy (to dark matter or something else) must have a good fit to the data. This is in agreement with \cite{Shafieloo09a}. However larger reconstructed $Om(z)$ at high redshifts (despite of the fact that the quality of the data is poor in this range) seems to be related to the apparent extra brightness of the supernovae data (in comparison with $\Lambda$CDM model) at $z > 1$ as reported and studied in~\cite{arman_leandros}. The second test $\mathscr{L}(z)$ relies on second derivatives of distance data, but is surprisingly useful even with present data; here, deviations from flat $\Lambda$CDM are seen with mild significance. It is interesting to note the close resemblance of the curves we have reconstructed here in a model independent way, to those same functions in void models~\cite{FLSC}.  

\item Test for the sign of curvature: $\mathscr{O}_k(z)D(z)^2$ is a test for flatness, which is independent of dark energy. It requires two independent data sets, and seems most useful at the moment for $z\lesssim1$. For $z\gtrsim1$ the errors on SNIa distances dominate. There is no evidence for deviation from flatness, though this is not surprising because it relies on Hubble rate data. 

\item Copernican test for homogeneity: $\mathscr{O}_k(z)$ is a model independent test for the FLRW models themselves, which is independent of dark energy or theory of gravity. It requires two independent data sets, and is very promising for $z\gtrsim1$. Again, because it relies on Hubble rate data, there is no evidence for deviations from FLRW, nor indeed from flatness. Note that for $0.8\lesssim z\lesssim1.6$ the data shows a mild trend towards closed models. This is likely of no significance. 
\end{itemize}

{It is early days for tests as sensitive as these because they require non-parametric construction, and we are still dealing with only a few hundred data points. We can easily envisage reaching thousands of data points in the next few years at which point these tests will come into their own as foundational tests for the concordance model. If the concordance model is correct then these tests will provide a useful way to quantify how confident we are in a more absolute way than we can at present. If, on the other hand, we find deviations for any of these tests, then they will help point us in other directions. In particular, if the Copernican tests $\mathscr{O}_k(z)$ and $\mathscr{C}(z)$ fail then we have a serious problem for cosmology, and the standard FLRW models. But with that would come a strong indication that dark energy cannot be due to modified field equations on either side; rather, we would know that there is something wrong with the spacetime metric we are using or that we are not describing the universe in the mathematically correct way.}

\section*{Acknowledgments}

AS acknowledge the support of the EU FP6 Marie Curie Research and
Training Network ``UniverseNet" (MRTN-CT-2006-035863). CC is supported by the NRF (South Africa). CC would like to thank the Department of Astrophysics at the University of Oxford for hospitality while part of this work was undertaken.

\end{document}